# A Case Study of Promoting Informal Inferential Reasoning in Learning Sampling Distribution for High School Students

**Geovani Debby Setyani**
Universitas Sanata Dharma, Yogyakarta, Indonesia
*geovanidebbys@gmail.com*

**Yosep Dwi Kristanto**
Universitas Sanata Dharma, Yogyakarta, Indonesia
*yosepdwikristanto@usd.ac.id*

**ABSTRAK**

Penarikan kesimpulan dari data adalah keterampilan penting bagi siswa untuk memahami kehidupan sehari-hari mereka, sehingga distribusi sampling yang menjadi topik utama statistika inferensial diperlukan oleh siswa. Akan tetapi, masih sedikit yang diketahui tentang bagaimana cara mengajarkan topik ini kepada siswa sekolah menengah atas, terutama dalam konteks Indonesia. Oleh karena itu, penelitian ini mendemonstrasikan eksperimen pengajaran dengan tujuan untuk mengembangkan penalaran inferensial informal siswa dalam memahami distribusi sampling, serta mendeskripsikan persepsi siswa terhadap eksperimen pengajaran tersebut. Subjek dalam penelitian ini adalah tiga siswa kelas 11 dari salah satu sekolah swasta di Yogyakarta jurusan matematika dan ilmu pengetahuan alam. Metode pengumpulan data yang digunakan adalah observasi langsung terhadap proses pembelajaran distribusi sampling, wawancara, dan dokumentasi. Penelitian ini menemukan bahwa penalaran inferensial informal dengan pembelajaran berbasis masalah menggunakan masalah kontekstual dan data yang riil dapat memfasilitasi siswa memahami topik distribusi sampling, dan siswa memberikan tanggapan positif terhadap pengalaman belajar mereka.

## INTRODUCTION

Constructing inference from data is an important skill for students to understand their everyday life. Therefore, it is necessary for students to study concepts and procedures in drawing statistical inference. The inferential statistics provides formal methods to statistically infer the populations' characteristics based on the sample data. Drawing conclusions beyond the sample data which considering the sample variation is one of the main themes of the inferential statistics (Triola, 2018). The remaining theme involves an investigation of whether a pattern in data can be attributed to a real effect (Garfield et al., 2008). Both of the main themes in inferential statistics are the basis of statistics (Pratt & Ainley, 2008) and considered as stepping stones for students in working with data to solve the problems they encountered.

The central concept in the inferential statistics is a sampling distribution. The sampling distribution is the distribution of a statistics for all possible samples of a certain size from the population (Sudjana, 2005; McClave, Benson, & Sincich, 2011; Stockburger, 2011). With sampling distribution, it is possible to move beyond the known sample data and draw a conclusion with regard to its population statistics. Drawing such conclusion is plausible since, by the definition, the sampling distribution includes the statistics from all possible sample outcomes so that it can be used to estimate the probability of any particular sample outcome, a central process in inferential statistics. In general terms, sampling distribution links a known sample data to its population statistics to draw conclusions regarding populations' characteristics based on the sample data.

Aforementioned rationale gives insight on the importance of the inferential statistics and the concept of sampling distribution. Therefore, it is necessary to teach the distribution sampling to high school students. Unfortunately, the topic is not covered explicitly in Indonesian national curriculum (Ministry of Education and Culture of the Republic of Indonesia, 2018). Mathematics curriculum involves inferential statistics topics for the twelve-grade majoring in mathematics and natural science, namely binomial and normal distribution. It is nevertheless not clearly stated the topic of sampling distribution. Therefore, the study with regard to teaching experiments of sampling distribution for high school students will provide constructive nuance for the Indonesian education system.

The sampling distribution is a crucial topic to be learned by students, yet it is difficult to teach. Many scholars have been proposed several methods in teaching the distribution sampling, one of them is using simulation. Even though this method is promising (see, Lane, 2015; Turner & Dabney, 2015), the simulation method has to be designed and implemented carefully. Watkins, Bargagliotti, & Franklin (2014) show that this method leads students to misunderstand about the sampling distributions. The broader teaching approach is potential in making students learn sampling distribution concept. It is the approach that let the students begin their learning process of sampling distribution by using their informal ideas. Using students' informal ideas about a concept makes the concept more accessible for the students (Zieffler, Garfield, Delmas, & Reading, 2008). This notion of promoting students' informal ideas about statistical inference is described as informal inferential reasoning or informal statistical inference.

Many scholars have attempted to describe informal inferential reasoning. Pfannkuch (2006) define informal inferential reasoning as "the drawing of conclusions from data that is based mainly on looking at, comparing, and reasoning from distributions of data." Zieffler et al. (2008) suggested the term informal inferential reasoning as "the way in which students use their informal statistical knowledge to make arguments to support inferences about unknown populations based on observed samples." Makar and Rubin (2009) identified three main features which characterize informal inferential reasoning: (1) generalization beyond the data, (2) the use of data to back up this generalization, and (3) the employment of probabilistic language (statement of uncertainty) in describing this generalization.

In supporting students to utilize their informal inferential reasoning, a learning environment should be designed in a supportive manner. First, students should be facilitated to work with a complex problem (Makar, 2014). Problem-based learning is one of learning methods which is fit to the learning environment criterion. Problem-based learning makes students learn through facilitated problem solving (Hmelo-Silver, 2004). Second, students should be facilitated to encounter data with authentic context (Makar, 2014). The authentic context emerges in the real data. Using the real data which interest students will engage them in thinking about relevant statistical concepts derived from the data (Aliaga et al., 2010; Pfannkuch, 2011). Therefore, problem-based learning which employs real data in facilitating students' learning about a statistical concept has potential in supporting students' informal inferential reasoning.

Given the importance of sampling distributions for high school students and the potential of problem-based learning, which utilizing real data as authentic context for the problem, in supporting students' informal inferential reasoning, the aim of the present study is two-fold. First, the present study provides a detail analysis on the emergence of students' informal inferential reasoning in the designed teaching experiments. Second, the present study discuss students' perceptions toward their experience during the teaching experiments.

## METHOD

The present study employed a qualitative descriptive approach, namely case study to achieve the research aims. The case study illustrated the constructivist teaching experiment (Steffe & Thompson, 2000) that we designed and implemented to support students' informal inferential reasoning in the understanding sampling distribution of means.

### Data Collection Method

The data used along with the acquisition techniques in this study are observations, interviews, and documentation. We conducted observations by looking at how subjects capture the topic and understand the sampling distribution by a direct guide from the first author who also as a teacher in the teaching experiment.

Interviews were used to determine the participants' perception in learning sampling distribution with their informal inferential reasoning. The interviews were semi-structured interviews, through which we can dig up information based on the answers from the subjects, not just following the guidelines of interviews that have been made previously. The questions in the interview were prepared to ask about and know things which were not obtained during observations. It was also used to facilitate us in getting information about the perceptions and responses of the participants during the teaching experiment.

Documentation was a complement of observation and interview methods in qualitative research carried out by the authors during the study. We used an audio-visual recorder to record the learning process whereas an audio recorder was used during the interviews. These recording devices were useful in collecting data, especially in explaining descriptions of various situations and behaviour of the subject under study. Before recording, we had asked permission to all participants and explained the confidentiality of their data. This complied the research ethics of Universitas Sanata Dharma.

### Teaching Experiment Participants

The participants in the teaching experiment were three 11th grader students from one of the private high schools in Yogyakarta majoring in mathematics and natural science. The participants were selected based on their inexperience in studying the sampling distribution. Additionally, a criterion for selection was based on previous students' learning achievement in mathematics. As a result, the participants consisted of three students who were from three different levels of mathematics learning achievement, namely high, medium, and low.

### Teaching Experiment Setting

The learning process carried out in the teaching experiment started from the initial topics, namely sample and population, sample and population mean, and sampling technique, to sampling distribution. The teaching experiment was conducted in five phases which spanned in three meetings. The teaching experiment structure is shown in Figure 1.

| Understanding sample dan population. | Understanding statistics dan parameter, especially mean | Understanding representative sampling of the population. | Understanding sampling techniques. | Understanding sampling distribution of mean. |
|---|---|---|---|---|
| Analyzing how survey institutions conduct quick counts. | Analyzing the average ballot needed by each polling station. | Analyzing sampling in a quick count so that it could predict the results of real count. | Analyzing sampling techniques for different cases that occurred in high schools. | Analyzing how good the samples to describe the duration of the Korean drama duration in the first episode. |
| Quick Count | | | Student's Identity | Korean Drama |
| Phase 1 | Phase 2 | Phase 3 | Phase 4 | Phase 5 |
| Meeting 1 | | Meeting 2 | | Meeting 3 |

**Figure 1**. The structure of teaching experiment

We used problem-based learning to help students understand the topics. The problems used in the learning process were authentic-context problems with real data. The context in the problems were quick count, students' identity, and Korean drama movies. The context was designed following the students' interest and experience. Hence, before carrying out the teaching experiment, the first author conducted interviews to find out the background of the students. In addition, learning scenarios were used to help and guide the authors in conducting learning process.

**Data Analysis**

The data analysis involved four stages, namely collecting the data, reducing the data, presenting the data, and drawing conclusions. Data collection was done through observation and interviews. At this stage, the data that has been collected is changed to transcripts by simplifying the information collected in an easily understood form of writing. After that, the data was selected according to the focus of this research to make it easier for the authors to categorize the collected data. Reducing data means summarizing, choosing key things, focusing on important things, looking for themes and patterns, and discarding things that are not relevant in the present study. Thus, the overall picture would be more clear and make it easier for the authors to conduct further data collection, and look for it as needed (Merriam, 2009). The data that has been summarized were then interpreted and explained to describe the process of students understanding the sampling distribution with their informal inferential reasoning. The data was presented in the form of narrative description. After that, we conclude from the results of data analysis that has been done. Drawing conclusions in qualitative research might answer the research questions. The conclusions in qualitative research are expected to be new findings that have never before existed. These findings could be in the form of a description of an object that was previously still unclear so that it becomes clearer after being investigated (Merriam, 2009).

**RESULTS AND DISCUSSION**

We describe the emergence of students' informal inferential reasoning in learning sampling distribution and their perceptions toward the teaching experiment in the result section. Then, the interpretation of the results is presented in the discussion section.

**Results**

From the learning process and interviews, learning outcomes and student perceptions were identified based on the elements of informal inferential reasoning as revealed by Makar and Rubin (2009) as follows.
1. Generalization beyond the data

In phase 1, students were asked to analyze how the survey institution knew the average ballots for each polling station, students had thought to use a sample to predict the average population of ballots for each polling station. This indicated that the sample's mean can be used to *estimate* the population's mean.

In phase 3, students were asked to analyze how there could be predictions of survey institutions that missed the results of the population. Students answered that it was because the samples taken by each institution were different. This indicated that students understood the *variation* between samples in a population. Figure 2 represents the student's understanding with regard to the variation between samples.

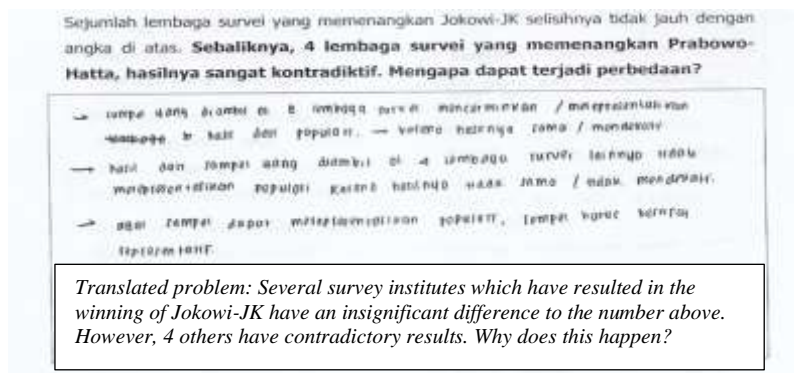

*Translated problem: Several survey institutes which have resulted in the winning of Jokowi-JK have an insignificant difference to the number above. However, 4 others have contradictory results. Why does this happen?*

**Figure 2.** Student's answer to quick count problem

In phase 4, students analyzed sampling techniques for four different cases. Each case had a population of 500. In this case, students can use an appropriate *sampling technique* to predict the populations' statistics.

In phase 5, students analyzed the mean of duration of the first episodes of Korean dramas that were popular and provided by seven different Korean dramas. Indeed, the students initially assumed that the population was the seven dramas so they choose to calculate the population's mean of the seven dramas. However, the first author emphasized that outside of the seven Korean dramas there were still many Korean dramas so what we would know was the population's mean of all Korean dramas. The seven Korean dramas here were used as a way to help facilitate sampling and as a substitute for the actual population used to view the most representative samples. So students choose to use a sample of 3.

2. Data as evidence

In phase 4, students analyzed the mean of duration of the first episode of Korean dramas that were popular and provided by seven different Korean dramas. Students used the seven Korean dramas as evidence by calculating the population's mean and sample differences.

3. The employment of probabilistic language (statement of uncertainty) in describing this generalization

During the learning process, students used the term sample and population because they already knew it beforehand.

According to Zieffler, Garfield, Delmas, & Reading (2008), there are several types of assignments that have been used in several studies and one of them is an assignment to consider two models or statements that are more likely to be true. This assignment could be found in phase 4 when students analyzed the average duration of the first episodes of Korean dramas that are popular and are provided by seven different Korean dramas. To find out which sample best represents the population or know how good the sample is taken, students calculated the difference in average population and average samples taken, then compared average samples to consider samples that are more likely to be true.

Prodromou (2013) emphasized that making conclusions informally gives students a feeling about the power of statistics to make judgments and reasonable decisions about data taken from real-world contexts. Every case/problem provided by the authors were in accordance with the students because cases/problems are deliberately taken from the real-world context. This was also realized by students. In interviews, students stated that the cases/problems provided, which were taken from real-world contexts, felt closer to them so that they found it easier to understand.

In this study, interviews were conducted on three students who had carried out the learning process of sampling distribution with informal inferential reasoning. The interview aims to find out the perceptions of students on learning distribution sampling with informal inferential reasoning. The results of this interview would be discussed based on the following aspects.

1. Overall Learning Process

    The students stated that the learning process was quite enjoyable. According to them, the learning process ran well in the right order, starting from the problems and then analyzing and discussing together to determine the right solution.

    Students emphasized the issues raised in cases. In the learning process, the authors used problem-based learning so that the authors raised problems to provoke students' analysis. Students explained that these problems made topics easier to understand because the problems could be found around them. In addition, learning is also more conducive and easier to understand because only three students took part in learning so that the teacher could interact more deeply and directed in joint discussions.

    S1 : In my opinion, the learning process was fun and it became easier than the learning in school. It was just three of us so the teacher could make more interaction with three of us. The problem also made understanding the topics easier, such as KPU, TPS.

    S2 : The learning process was already good and so was the plot, from analyzing the problems to finding the solutions. In my opinion, the process was successful because the problems and the solutions were clear.

    S3 : It was easy to understand the topics because the teacher gave the problems from our interests.

2. Students' Perceptions of the Learning Process

    After carrying out learning distribution sampling with informal inferential reasoning, the students mentioned the things that were obtained from the learning which included sample and population, mean of the sample and population, representative sample, and how good the sample was taken based on the average sample difference and average population.

    S1 : I could know about sample and population, also the sample mean, the population mean, and a sample which best represents a population, based on the difference between the sample mean and the population mean.

    S2 : I could know about populations, sample, how I can get samples which represent populations the most, and also quick count. I just found those problems that came from daily life.

        **S3** : I could know about the sample and population's mean and the population mean, and the difference between them to get samples which represent populations the most.

3. Students' Feelings towards the Learning Process

    Students stated that during learning distribution sampling learning with informal inferential reasoning, students felt that learning took place pleasantly and they did not feel burdened at all on the first day, the longer it was made the learning more difficult because they had to think harder, but it paid off because they could understand and the problems raised made it easier to understand.

        **S1** : I was not burdened by the learning.
        **S2** : It was fun.
        **S3** : The first day was enjoyable because it was still easy. The second day became more difficult, we should think harder and I felt sleepy. The last day was hard. But it was all good because I could understand.

4. The Most Interesting Topic about Learning

    In the learning process of distribution sampling with informal inferential reasoning, the problems raised were deliberately taken from the things that were around the students. Therefore, these problems became some of the things that were quite interesting for students, especially in the case of the mean of duration of Korean drama movies. It became something new for them to find mathematical problems. In addition, quick counts were also quite interesting because they did not know how the quick count results were obtained. After this learning, the students realized that the quick count conducted by survey institutions had to use the right sample so that the results were accurate.

        **S1** : The most interesting topic was the Korean drama case because it was close to high school students' life. We usually watch Korean drama but we never knew it could be a case.
        **S2** : The most interesting topic was quick counts in Indonesia, how we count. We usually just watch it on TV and don't know that the quick count is from a sample of a population. I could know how to calculate it so the results could be this accurate.
        **S3** : The most interesting topic was sampling techniques.

5. The Easiest and The Hardest Topic

    Based on the interviews, the three students stated that the easiest topic was understanding the sample and population, also the average sample and average population. They said it was because the thinking process was still quite easy and did not require a high level of analysis. On the other hand, the topic in phase 5 was the most difficult, namely knowing how well a sample is. This was due to the possibility of huge samples to be registered in Microsoft Excel. In addition, analyzing the sampling techniques for cases presented was also considered difficult because more analysis was needed among several of these cases.

        **S1** : The easiest topic was quick counts because we just had to know which one was sample and population. The hardest topic was finding how good the sample was because it was hard to analyze.
        **S2** : The easiest topic was the sample mean and the population mean. The hardest topic was sampling techniques because we should know the conditions for each case.

    **S3** : The easiest topic was sample and population. The hardest topic was finding how good the sample was because it was hard to analyze. It was also because there were so many samples.

6. Students' Perception of the Learning Method

During the sampling distribution learning process, we facilitate students' informal inferential reasoning with problem-based learning to help them to understand the sampling distribution. Provoking questions were also used along with the contextual problems and real data to arouse students' curiosity. According to students, it helped them to understand the topic because of the problems taken from around their real lives.

    **S1** : It was good because we could understand easily and the problems were close to us.
    **S2** : The method helped us by analyzing the cases we could understand.
    **S3** : It helped because the problems were around us so we could relate to it.

By this approach, students stated that nothing needed to be improved because they felt the learning was very enjoyable, they emphasized the problems taken from around their lives. However, in phase 5 when students began to take samples and calculated the sample mean, the students did not take samples randomly as they had studied in the previous phase. The students chose their samples based on Korean dramas they knew. One of the students stated it would be better, before taking samples, students were told to use the lottery method. We deliberately did not tell to use random sampling because we hoped that the students could realize and analyze the sampling method themselves.

    **S1** : No, [There is no need for improvements]
    **S2** : To get the sample [in phase 5] it would be better if the teacher explained how we take the sample.
    **S3** : I think no because I hope there are many teachers like this. The problems used came from around us.

7. The Importance of Sampling Distribution

After carrying out learning distribution sampling with informal inferential reasoning, students stated that studying the sampling distribution was quite pleasant because they could know the importance of sampling to get accurate results and also very useful in dealing with news to avoid hoaxes. Besides, there were many things around the students such as height mean and blood types which could be identified. However, students felt that the sampling distribution has not been used yet, but maybe later when working but it also depends on the work as well.

    **S1** : For now, it was not useful. Maybe in the future, when we work, we can use it. Also, it depends on our job. But it is very useful to respond to the news.
    **S2** : The problems used can be found around us so that we can know how to solve it. We should know how to get the best sample by finding the difference.
    **S3** : It was important. Calculating means becomes easier by not having to use the population.

8. The Students' Perception of Their Fellow Students' Readiness in Learning Sampling Distribution

After carrying out the sampling distribution learning with problem-based learning, students have understood the topic provided, namely sample, population, the sample mean, the population mean, sampling techniques, and sampling distribution. According to them, other high school students majoring in mathematics and science, such as their classmates, could understand this sampling distribution with the same terms of the approach used namely informal inferential reasoning and problem-based learning with problems taken from around high school students' life and discuss each other to determine the solution of the problems. They mentioned that if the reasoning used was formal reasoning like in school, maybe this topic would be more difficult to understand.

> S1 : I think they can [be applied more widely] because the learning method used made it easier to understand. It can be hard if it had used formal methods like in school.
> S2 : By using the same method, they can understand because discussions make it easier in Stece [Stella Duce 1 High School Yogyakarta].
> S3 : They can understand if the method used is the same as ours. If the teacher uses formal methods, it may be hard.

**Discussion**

Based on the learning process of distribution sampling with informal inferential reasoning and the interviews that had been conducted, there were several aspects that we found to be the key aspects in developing $11^{th}$-grader in understanding the sampling distribution, namely informal inferential reasoning, problem-based learning, and the use of real data.

Informal inferential reasoning is an informal method or process in drawing conclusions and generalizing from a group of data for a wider range of data. The essence of informal inferential reasoning is an approach that is carried out informally, that is, not a teacher who gives theories or hypotheses to be tested. In this reasoning, students start from the problem and, with the ability and schema of existing knowledge, analyze and conclude which sample is better for representing a population. Informal inferential reasoning could be found to appear in the learning process, as expressed in the results section, based on informal inferential reasoning frameworks by Makar and Rubin (2009), Zieffler, et al (2008), and Prodromou (2013).

Based on the learning process that has been carried out and the results of interviews, sampling distribution learning was easier to understand using informal inferential reasoning. Students said if the reasoning used in the learning process of the sampling distribution had been formal reasoning like in school, the sampling distribution topic might have been more difficult to understand. Students felt that the learning that has been done was enjoyable and they did not feel burdened at all with these topics because the learning took place more informally and not rigidly.

Problem-based learning is a learning approach that aimed to enable students to solve a problem to develop their knowledge, develop inquiry and high-level thinking skills, and improve self-reliance and self-confidence. Problem-based learning was conducted to help students develop thinking, problem-solving, and intellectual skills. Cultivating students' knowledge, which is then used to share knowledge with friends, could increase self-confidence in the learning. In the learning process of the sampling distribution, each student was allowed to analyze problems and then discuss them with others. This discussion was very helpful for students to jointly build knowledge. This was also supported by the number of students who took part in the learning, there were only three students so that the teacher could easily reach every student. The learning was also conducted in no hurry to provide opportunities for students to think more critically and try to find the basis of their arguments and facts that supported the reasons to make conclusions.

During the learning process, there was one student who only had a little discussion and just listened to two other students conveying the results of their analysis. When the student was asked, she found it difficult to explain again what had been learned. After the teacher and other students helped the student by explaining again, the student was able to understand. It indicated that students were expected to be more active in analyzing so that they could build their knowledge and then discuss it together.

Problem-based learning was included in active learning to invite students to learn actively, to participate in the learning process. The students, who only received topics from the teacher and assumed that just memorizing the topic was enough, must be changed to find knowledge actively so that there could be an increase in understanding. This learning model was oriented toward building students' knowledge independently. In problem-based learning, the focus of learning was on the problem chosen so that students not only learned concepts related to the problems but also how to solve the problem. Therefore, students also gained learning experiences related to problem-solving skills. Students who learned to solve a problem would apply the knowledge they had or tried to find out the knowledge needed. Learning could be more meaningful and expanded when students were faced with situations where the concept was applied.

Nowadays, students sometimes are too lazy to analyze and tend just to answer a question by directly answering or searching from other sources without having the intention to try to analyze or express their own opinions. If this situation continues, the students in the study would have difficulty applying the knowledge gained in the classroom in real life. Therefore, problem-based learning could be a solution to encourage students to think critically and work rather than memorizing.

Based on research conducted by Farida and Kusmanto (2014), after the application of problem-based learning, there was an increase in interest and mathematical achievement by meeting the minimum completeness criteria. In addition, research conducted by Abdullah (2016) indicated that the application of problem-based learning could improve the quality of learners' mathematics learning process seen from the increase in student learning outcomes tests. The similar result is also found on Sahrudin and Trisnawati (2018). These indicated that problem-based learning increase students' interest and learning outcomes which indicated that the model made the learning process better.

Problem-based learning could not be separated from contextual problems taken from everyday life. The authors have interviewed the students regarding their backgrounds to find out what things could be raised as problems in learning this sampling distribution. As explained in the previous section, Prodromou (2013) emphasized that making conclusions informally gave students a feeling about the power of statistics to make judgments and sensible decisions about data taken from real-world contexts. The use of contextual problems was very helpful for students to understand the distribution of sampling with informal inferential reasoning because the real-world context was closed to students.

Contextual problems that were the focus of learning could motivate students to solve the problems because the problems created disequilibrium between concepts in the cognitive schemes of students and the contexts. It encouraged curiosity so that it raised many questions about why and how for the problem. These questions were the motivation for students to learn. From this explanation, it could be seen that contextual problems could encourage students to have initiatives to build their knowledge.

Contextual problems encouraged students to analyze a phenomenon by the emergence of questions. Conversation and collaboration could help in the process of answering questions, which are carried out in discussions. Informal discussions could create collaboration. Informal inferential reasoning was very helpful in solving contextual problems because the problems emerge from the students themselves so they could collaborate with each other. This finding aligns Rahmadonna & Fitriyani (2011), that participants' learning motivation increases with the application of mathematics learning with a contextual approach. This indicated that students have more motivation to solve a problem.

Motivation is very influential in building curiosity and willingness to learn. With the problems that came from around the students, students had more willingness to study and analyze these problems. The learning process was centered on learners, where they could analyze problems which then provided the learning process experience that encouraged students to conduct research, integrate theory, and apply the knowledge as well as skills they have in providing solutions to problems.

In addition to contextual issues, the cases raised also included real data, that existed and was not made up. The real data were the results of the 2014 quick count and Korean dramas. These data did exist and we took from trusted sources. We raised real data so that the problems presented to students were felt more real and students had a sense of trust in the data provided.

Real data that the we used helped students to understand the main issues in real terms. Real data could help students learn to identify the root of the problem to improve students' critical thinking skills that are very useful in life. It could encourage students to realize that the problems were in accordance with real conditions and were no longer theories so that problems in the application of a concept could be found during learning.

Research conducted by Partono (2009) found that students 'learning achievements with contextual learning models were better than students' learning achievements with direct learning models. Based on the results of these studies, learning based on contextual problems helps students to relate learning topic to real-world situations and encourage students to make connections between the knowledge they have and their application in daily life to produce more meaningful learning processes and outcomes that valuable for students.

**CONCLUSION**

The present study aimed to provide insight on how informal inferential reasoning and problem-based learning with real data and authentic context help students in understanding sampling distribution concept, as well as describe students' perceptions toward the teaching experiment. Additionally, the present study provides learning scenarios which can be utilized by the researcher and practitioner to design the learning process in the topic of a sampling distribution. The learning scenario is an empirical-based learning design that supports students' informal inferential reasoning in the understanding sampling distribution. The present study found that students have positive perceptions toward the designed learning scenario.

Based on the findings, we suggest future researchers conduct similar research in a larger and heterogeneous class. It aims to deepen the discussion in informal inferential studies so that they can be applied to a larger class. In this study, contextual problems and real data had a positive effect on learning the sampling distribution. Contextual issues and real data helped students understand the topic more quickly. Therefore, we suggest teachers use contextual problems in helping students understand a statistical concept in particular and other topics in general.

**ACKNOWLEDGEMENT**

The authors would like to acknowledge the students who participated in this study and the two anonymous reviewers who gave constructive suggestions to improve the earlier version of this manuscript.

# A Case Study of Promoting Informal Inferential Reasoning in Learning Sampling Distribution for High School Students

**Geovani Debby Setyani**
Universitas Sanata Dharma, Yogyakarta, Indonesia
*geovanidebbys@gmail.com*

**Yosep Dwi Kristanto**
Universitas Sanata Dharma, Yogyakarta, Indonesia
*yosepdwikristanto@usd.ac.id*

**ABSTRACT**

Drawing inference from data is an important skill for students to understand their everyday life, so that the sampling distribution as a central topic in statistical inference is necessary to be learned by the students. However, little is known about how to teach the topic for high school students, especially in Indonesian context. Therefore, the present study provides a teaching experiment to support the students' informal inferential reasoning in understanding the sampling distribution, as well as the students' perceptions toward the teaching experiment. The subjects in the present study were three 11th-grader of one private school in Yogyakarta majoring in mathematics and natural science. The method of data collection was direct observation of sampling distribution learning process, interviews, and documentation. The present study found that that informal inferential reasoning with problem-based learning using contextual problems and real data could support the students to understand the sampling distribution, and they also gave positive responses about their learning experience.

***Keywords:*** *Informal Statistical Inference, Sampling Distribution, Problem-Based Learning, Real Data, Statistics Education Research*